\documentclass[11pt,a4paper,english,nofootinbib]{revtex4}
\usepackage{lmodern}

\usepackage[T1]{fontenc}
\usepackage[latin9]{inputenc}
\usepackage{color}
\usepackage{babel}
\usepackage{amsmath}
\usepackage{amssymb}
\usepackage{esint}
\usepackage[unicode=true,pdfusetitle,
 bookmarks=true,bookmarksnumbered=false,bookmarksopen=false,
 breaklinks=false,pdfborder={0 0 1},backref=false,colorlinks=false]
 {hyperref}

\makeatletter

\pdfpageheight\paperheight
\pdfpagewidth\paperwidth

\@ifundefined{textcolor}{}
{%
 \definecolor{BLACK}{gray}{0}
 \definecolor{WHITE}{gray}{1}
 \definecolor{RED}{rgb}{1,0,0}
 \definecolor{GREEN}{rgb}{0,1,0}
 \definecolor{BLUE}{rgb}{0,0,1}
 \definecolor{CYAN}{cmyk}{1,0,0,0}
 \definecolor{MAGENTA}{cmyk}{0,1,0,0}
 \definecolor{YELLOW}{cmyk}{0,0,1,0}
}

\usepackage{latexsym}\usepackage{bm}

\makeatother

\definecolor{green}{RGB}{0, 180, 0}
\definecolor{cyan}{RGB}{0, 180, 180}
\definecolor{yellow}{RGB}{211,211,0}

\makeatother

\begin{document}

\title{Near Horizon Geometry of Warped Black Holes in Generalized Minimal Massive Gravity}

\author{Mohammad Reza Setare}
\email{rezakord@ipm.ir}

\affiliation{{Department of Science,\\
 Campus of Bijar, University of Kurdistan, Bijar, Iran\\
 Research Institute for Astronomy and Astrophysics of Maragha (RIAAM),
P.O. Box 55134-441, Maragha, Iran }}

\author{Hamed Adami}
\email{hamed.adami@yahoo.com}

\affiliation{{Research Institute for Astronomy and Astrophysics of Maragha (RIAAM),
P.O. Box 55134-441, Maragha, Iran }}

\date{\today}

\begin{abstract}
We consider spacelike warped AdS$_{3}$ black hole metric in Boyer-Lindquist coordinate system. We present a coordinates transformation so that it maps metric in Boyer-Lindquist coordinates to the one in Gaussian null coordinates. Then we introduce new fall-off conditions near the horizon of non-extremal warped black holes. We study the near horizon symmetry algebra of such solutions in the context of Generalized minimal massive gravity. Similar to the black flower solutions, also we obtain the Heisenberg algebra as the near horizon symmetry algebra of the warped black flower solutions. We show that the vacuum state and all descendants of the vacuum have the same energy. Thus these zero energy excitations on the horizon appear as soft hairs on the warped black hole.
\end{abstract}

\maketitle

\section{Introduction}\label{S.I}
It is well known that the Chern-Simons-like theories of gravity (CSLTG) in $(2+1)$-dimension \cite{1} (e.g. TMG \cite{1'}, NMG \cite{2''}, MMG \cite{3''}, Zwei-dreibein gravity (ZDG)\cite{4'}, GMMG \cite{5'}, etc), exhibit local physical degrees
of freedom, but for these theories, different boundary conditions can lead to
completely different boundary theories. For matter-free Einstein-Hilbert gravity, the behavior of the three-dimensional metric at spatial infinity
is given by the Brown-Henneaux boundary conditions \cite{12}. But in the
presence of matter, these boundary conditions can be modified \cite{13}. This modification can be occurs in Topological massive gravity \cite{14} and even in pure Einstein-Hilbert gravity \cite{15}.\\
The realization of the existence of three-dimensional (3D) black holes deepened our understanding of 3D gravity. In this context also an important role is played by the notion of asymptotic symmetry. The near horizon symmetries of the black hole solutions in three dimensions are related with the BMS algebra \cite{50'}. The BMS symmetry algebra in $n$ space-time dimension consists
 of the semi-direct sum of the conformal Killing vectors of a $(n-2)$-dimension sphere acting on the ideal of infinitesimal
 supertranslations \cite{40',50''}. Recently, motivated in great part by the recent work of Hawking, Perry and Strominger \cite{26a} (for another related works see \cite{7c}), it has appeared that a new way to approach the information paradox for black holes lies in a careful analysis of near horizon symmetries and the existence (in 4 dimensions) of an infinite-dimensional asymptotic symmetry group, the BMS group. Roughly, it is hoped that the fact the latter yields an infinite number of (asymptotically) conserved charges would allow to store information about the collapse into a black hole and to evade the no-hair theorem, because different final black hole states, though diffeomorphic, would carry different BMS charges.
 \\ Recently Donnay et al \cite{500}, have shown that the asymptotic symmetries close to the horizon of the nonextremal black hole solution of the three-dimensional Einstein gravity in the presence of
a negative cosmological term, are generated
by an extension of supertranslations .  The authors of \cite{500} have shown that for a special choice of boundary conditions, the
near region to the horizon of a stationary black hole
present a generalization of supertranslation, including a
semidirect sum with superrotations, represented by Virasoro
algebra. More recently, we have studied the behaviors and algebras of the symmetries and conserved charges near the horizon of the non-extremal black holes in the context of the so-called Generalized Minimal Massive Gravity \cite{140}, proposed in Ref. \cite{5'}. In the other hand the authors of \cite{9} have considered the black flower solution of the Einstein
equations in $3D$ \cite{16}, then have proposed
a new set of boundary conditions, which leads to a very simple near horizon symmetry algebra,
the Heisenberg algebra. In another paper we have studied this near horizon symmetry in the framework of Chern-Simons-like theories of gravity \cite{3}. In another term, similar to the near horizon symmetry algebra of the black flower solutions of the Einstein gravity in the presence of negative cosmological constant, we have found an algebra consists
of two $U(1)$ current algebras, but instead with levels $\pm\frac{k}{2}$, the level of our algebra is given by $\mp \frac{k}{2} \left( \sigma \pm \frac{1}{\mu l}+ \frac{\alpha H}{\mu} + \frac{F}{m^{2}} \right)$. The authors of \cite{9} have obtained a new horizon entropy formula, as $S=2\pi (J_{0}^{+}+J_{0}^{-})$ for black hole solutions in $3D$ Einstein gravity, where $J_{0}^{\pm}$ are zero mode charges of $U(1)$ current algebra near horizon. More recently Gonzalez et al, have obtained an analog of the above entropy formula for non-extermal Kerr black holes in 4-dimensions \cite{31a}.\\
In this paper we are going to study the near horizon symmetry of spacelike warped AdS$_{3}$ black hole solutions of GMMG. The authors of \cite{21b} have introduced spacelike stretched $AdS_3$ as a new vacuum of TMG, which could be a stable ground state of this model \cite{15'b,23b} (see also \cite{18'b,19'b,20'b}). More than this, another reason for interest to the warped $AdS$ spacetime, is that they emerge
in the near horizon geometry of extremal Kerr black holes and this fact is important in the context of Kerr/CFT \cite{21'b}. In the paper \cite{23b}, the correctness of the hypothesis formulated in \cite{21b} have been investigated. The authors of  \cite{23b} have obtained that the asymptotic symmetry of the spacelike stretched $AdS_3$  sector of TMG is a 2-dimensional conformal
symmetry with central charges. Recently warped $AdS_3$ black holes have been studied in generic higher derivatives gravity theories in $2+1$ dimensions \cite{6a}. According to earlier investigations on the warped $AdS_3$, the asymptotic symmetry group of these spacetimes instead of the Brown-Henneaux conformal symmetry group, is the semi-direct product of a Virasoro algebra and a $U(1)$ affine Kac-Moody algebra \cite{23'b,24'b,25'b,23b,16b}. It is the symmetry of warped conformal field theory (WCFT) in 2-dimension. The authors of \cite{6a} have reproduced the match between the Bekenstein-Hawking and WCFT entropies in the case of new massive gravity (NMG) \cite{2''}. \\
Here we consider spacelike warped AdS$_{3}$ black hole metric in Boyer-Lindquist coordinate system as a solution of GMMG. Then we introduce a coordinates transformation such that it maps the metric into the one in Gaussian null coordinates. Based on the near horizon behaviour of the obtained metric of non-extremal warped black holes, we propose a new and consistent set of fall-off conditions. Then we find the conserved charges of the new fall-off conditions. Similar to the near horizon symmetry algebra of the black flower solutions in Generalized minimal massive gravity \cite{3}, the near horizon symmetry algebra of the warped black flower consists of two $U(1)$ current algebras, with different levels. In this case, the levels are $\mp \frac{\zeta^{2} \nu^{2} c_{\pm}}{48}$. By changing the basis, finally we obtain the Heisenberg algebra as the near horizon symmetry algebra of the warped black flower solutions. Then we show that the formula $S=2\pi (J_{0}^{+}+J_{0}^{-})$ exactly works for warped black flower solutions. \footnote{In this paper we have denoted zero mode charges by $\mathcal{L}_{0}^{\pm}$}.
\section{Note added:}
 While finishing our work we became aware of \cite{1c} which studies soft hairy warped black hole entropy in topological massive gravity.
\section{Warped black holes}\label{S.II}
Consider spacelike warped AdS$_{3}$ black hole which is described by the following line-element \cite{1a}
\begin{equation}\label{1}
  ds^{2}= g_{tt} dt^{2} + 2 g_{t \theta} dt d \theta +g_{rr} dr^{2}+g_{\theta \theta} d \theta ^{2},
\end{equation}
with
\begin{equation}\label{2}
  \begin{split}
       & g_{tt}=l^{2}, \hspace{0.7 cm} g_{t \theta}= \frac{1}{2} l^{2} \left| \zeta \right| \left( r + \nu \sqrt{r_{+} r_{-}} \right), \hspace{0.7 cm} g_{rr}= \frac{l^{2}}{\zeta^{2} \nu^{2}\left( r - r_{+} \right)\left( r - r_{-} \right)}\\
       & g_{\theta \theta}=\frac{1}{4}l^{2} \zeta ^{2} r \left[ \left( 1 - \nu ^{2} \right) r + \nu ^{2} \left( r_{+} + r_{-} \right) + 2 \nu \sqrt{r_{+} r_{-}} \right],
  \end{split}
\end{equation}
where $t$, $r$, $\theta$ and $l$ are time-coordinate, radial-coordinate, angular-coordinate and AdS$_{3}$ space radius, respectively. For the spacelike warped AdS$_{3}$ black hole $r_{+}$ and $r_{-}$ are outer and inner horizons, respectively. The free parameters $\zeta$ and $\nu$ appeared in Eq.\eqref{2} allow us to keep contact with \cite{2a,3a,4a,5a}\footnote{For further discussion, see \cite{1a}.}. The spacetime described by metric \eqref{1} admits $SL(2,\mathbb{R})\times U(1)$ as isometry group. Therefore, one can write a symmetric-two tensor $S_{\mu \nu}$ as \cite{6a}
\begin{equation}\label{3}
  S_{\mu \nu}=a_{1} g_{\mu \nu} +a_{2} J_{\mu} J_{\nu},
\end{equation}
with
\begin{equation}\label{4}
  J=J^{\mu}\partial_{\mu}=\partial_{t}.
\end{equation}
It should be noted that $J_{\mu}$ is normalized as $J_{\mu} J^{\mu}=l^{2}$ and it satisfy the following equation
\begin{equation}\label{5}
  \nabla _{\mu} J_{\nu} = \frac{\left| \zeta \right|}{2l} \epsilon_{\mu \nu \lambda} J^{\lambda}
\end{equation}
where $\nabla _{\mu}$ is covariant derivative with respect to the Christoffel connection and $  \epsilon_{\mu \nu \lambda} = \sqrt{-g}  \varepsilon_{\mu \nu \lambda}$ is Levi-Civita tensor in 3D. Hence, the Ricci tensor can be written as
\begin{equation}\label{6}
  \mathcal{R}_{\mu \nu} = \frac{\zeta^{2}}{2l^{2}}\left( 1-2 \nu ^{2} \right) g_{\mu \nu} - \frac{\zeta^{2}}{l^{4}} \left( 1- \nu ^{2} \right) J_{\mu} J_{\nu},
\end{equation}
and the Ricci scalar is given by
\begin{equation}\label{7}
  \mathcal{R}= \frac{\zeta^{2}}{2l^{2}}\left( 1-4 \nu ^{2} \right).
\end{equation}
The authors in Ref.\cite{7a} showed that any line-element for which the Ricci tensor can be written as Eq.\eqref{6}, where the vector field $J$ is normalized to $l^{2}$ and satisfies Eq.\eqref{5}, is a solution of Generalized minimal massive gravity.
\section{Near horizon geometry}\label{S.III}
The near horizon geometry of three-dimensional black holes can be expressed using Gaussian null coordinates
\begin{equation}\label{8}
  ds^{2}= g_{vv}dv^{2}+2 dv d \rho + 2 g_{v \phi} dv d \phi +2 g_{\rho \phi} d\rho d\phi +g_{\phi \phi} d \phi ^{2},
\end{equation}
where $v$ represents the retarded time, $\rho$ is the radial distance to the horizon and $\phi$ is the angular coordinate. The metric
deviates to usual from in the $g_{\rho \phi}$ component,
but this can always be removed by a coordinates transformation. However, it is convenient to keep it as it is. The non-zero components of metric are demanded to obey the following fall-off conditions close to the horizon ($\rho=0$):
\begin{equation}\label{9}
\begin{split}
     & g_{vv}= -2 \kappa _{H} \rho + \mathcal{O} (\rho ^{2}), \hspace{0.7 cm} g_{v \phi}= A(\phi) \rho  + \mathcal{O} (\rho ^{2}), \hspace{0.7 cm} g_{\rho \phi}= B (\phi) + \mathcal{O} (\rho ^{2}), \\
     & g_{\phi \phi}= C(\phi)+ D(\phi) \rho + \mathcal{O} (\rho ^{2}),
\end{split}
\end{equation}
where $A(\phi)$, $B(\phi)$, $C(\phi)$ and $D(\phi)$ are arbitrary functions and the constant $\kappa _{H}$ corresponds to the black hole surface
gravity. We can relate the spacelike warped AdS$_{3}$ black hole metric in Boyer-Lindquist coordinates \eqref{2} to the one in Gaussian null coordinates by introducing a coordinates transformation as
\begin{equation}\label{10}
  \begin{split}
     dv= & dt+\frac{2}{l^{4}} g_{t \theta} g_{rr} dr \\
      \rho = & \frac{\left( r-r_{+}\right)}{\left| \zeta \right| \left( r_{+} + \nu \sqrt{r_{+} r_{-}} \right)} \\
      d \phi = & \frac{\nu dt}{\left( r_{+} + \nu \sqrt{r_{+} r_{-}} \right)}+ \frac{1}{2} \left| \zeta \right| \nu d \theta + \frac{\left| \zeta \right| \nu}{l^{4}} \left[ \frac{2 g_{t \theta}}{\left| \zeta \right| \left( r_{+} + \nu \sqrt{r_{+} r_{-}}\right)}-l^{2}\right] g_{rr} dr.
  \end{split}
\end{equation}
Using the above coordinates transformation, one can show that the line-element \eqref{1} can be rewritten as
\begin{equation}\label{11}
  \begin{split}
      \frac{ds^{2}}{l^{2}}= & -2\kappa_{H} \rho f(\rho) dv^{2}+2 dv d\rho +2 \left| \zeta \right| \nu ^{-1} \rho \left[ - \left( r_{+} + \nu \sqrt{r_{+} r_{-}}\right)+\nu^{2} (r_{+}-r_{-}) f(\rho) \right] dv d\phi \\
       & - \frac{\left| \zeta \right| \nu }{\kappa _{H}} (r_{+}-r_{-}) d \rho d\phi \\
       & + \nu ^{-2} \left[ \left( r_{+} + \nu \sqrt{r_{+} r_{-}}\right)^{2}+\frac{\zeta ^{2} \nu ^{2} \rho }{\kappa _{H}} \left( r_{+} + \nu \sqrt{r_{+} r_{-}}\right) (r_{+}-r_{-})-\frac{\zeta ^{2} \nu ^{4} \rho }{2 \kappa _{H}} (r_{+}-r_{-})^{2} f(\rho)  \right] d \phi ^{2},
  \end{split}
\end{equation}
where $\kappa_{H}$ is surface gravity of the warped black hole,
\begin{equation}\label{12}
  \kappa_{H}=\frac{\left| \zeta \right| \nu^{2}\left( r_{+}-r_{-} \right)}{2\left( r_{+} + \nu \sqrt{r_{+} r_{-}} \right)},
\end{equation}
and $f(\rho)$ as a function of $\rho$ is given by
\begin{equation}\label{13}
  f(\rho)= 1- \frac{\zeta^{2} (1-\nu^{2}) \rho}{2 \kappa_{H}}.
\end{equation}
It is clear that the line-element \eqref{11} obey the near horizon fall-off conditions. Since the vector $J=\partial_{t}$ can be written as
\begin{equation}\label{14}
  J=\partial_{v}+ \frac{\nu}{\left( r_{+} + \nu \sqrt{r_{+} r_{-}}\right)} \partial_{\phi},
\end{equation}
in Gaussian null coordinates then, one can show that the line-element \eqref{11} satisfies Eq.\eqref{6}. This ensures that the coordinates transformation \eqref{10} is well-defined because it maps warped black hole spacetime on itself.\\
Now we can claim that the behaviour of the warped black hole spacetime around a non-extremal horizon can be described by a near horizon metric in Gaussian null coordinates
\begin{equation}\label{15}
  \begin{split}
      \frac{ds^{2}}{l^{2}}= & -2\kappa_{H} \rho f(\rho) dv^{2}+2 dv d\rho +2 \left| \zeta \right| \nu ^{-1} \rho \left[ - \gamma (\phi)+\nu^{2} \beta (\phi) f(\rho) \right] dv d\phi \\
       & - \frac{\left| \zeta \right| \nu }{\kappa _{H}} \beta (\phi) d \rho d\phi + \nu ^{-2} \left[ \gamma(\phi)^{2}+\frac{\zeta ^{2} \nu ^{2} \rho }{\kappa _{H}} \gamma(\phi) \beta(\phi)-\frac{\zeta ^{2} \nu ^{4} \rho }{2 \kappa _{H}} \beta(\phi)^{2} f(\rho)  \right] d \phi ^{2},
  \end{split}
\end{equation}
where $\gamma(\phi)$ and $\beta(\phi)$ are arbitrary functions of $\phi$. One can show that the metric \eqref{15} with vector field
\begin{equation}\label{16}
  J= \frac{\left| \zeta \right| \nu^{2} \beta(\phi)}{2 \kappa_{H} \gamma(\phi)}\partial_{v}+\frac{\nu}{\gamma(\phi)}\partial_{\phi}.
\end{equation}
satisfies Eq.\eqref{6}. Also, it can be checked that the vector field $J$ (introduced in Eq.\eqref{16}) is normalized as $J_{\mu} J^{\mu}=l^{2}$ and it obeys Eq.\eqref{5}. Thus, the metric \eqref{15} is a solution of Generalized minimal massive gravity (we will discuss this later). The given geometry possesses an event horizon located at $\rho =0$ and also it is not spherically symmetric, thus the metric \eqref{15} generically
describes a "warped black flower". In the case of $\gamma =r_{+} + \nu \sqrt{r_{+} r_{-}}$ and $\beta=r_{+}-r_{-}$ the metric \eqref{15} reduces to the warped black hole metric.
\section{New fall-off conditions}\label{S.IV}
Based on the near horizon behaviour of the metric, we can propose a new set of fall-off conditions
\begin{equation}\label{17}
  \begin{split}
      \frac{ds^{2}}{l^{2}}= & \rho \left( a^{+} f_{+}+a^{-} f_{-} \right) dv^{2}+2 dv d\rho + \frac{2 \left| \zeta \right| \nu \mathcal{L}^{-}}{(a^{+}+a^{-})} d \rho d\phi  \\
      & +2 \left| \zeta \right| \nu ^{-1} \rho \left[ - \frac{1}{2} \left( \mathcal{L}^{+}+\mathcal{L}^{-}\right)+\frac{ \nu^{2} \mathcal{L}^{-}}{(a^{+}+a^{-})} \left( a^{+} f_{+}+a^{-} f_{-} \right) \right] dv d\phi \\
      + \nu ^{-2} &  \left[ \frac{1}{4} \left( \mathcal{L}^{+}+\mathcal{L}^{-}\right)^{2}-\frac{\zeta ^{2} \nu ^{2} \rho }{(a^{+}+a^{-})} \left( \mathcal{L}^{+}+\mathcal{L}^{-}\right) \mathcal{L}^{-}+\zeta ^{2} \nu ^{4} \rho \left(\frac{\mathcal{L}^{-}}{a^{+}+a^{-}}\right)^{2} \left( a^{+} f_{+}+a^{-} f_{-} \right)  \right] d \phi ^{2},
  \end{split}
\end{equation}
where $\mathcal{L}^{\pm}= \mathcal{L}^{\pm}(\phi)$ are arbitrary functions of $\phi$, $a^{\pm}$ are constant parameters and $f_{\pm}= f_{\pm}(\rho)$ are given as
\begin{equation}\label{18}
  f_{\pm}(\rho)= 1+ \frac{\zeta^{2} (1-\nu^{2}) \rho}{2 a^{\pm}}.
\end{equation}
In the particular case of $a^{\pm}=-\kappa_{H}$, $\mathcal{L}^{+}= 2 \gamma - \beta$ and $\mathcal{L}^{-}= \beta$ the line-element \eqref{17} will be reduced to the line-element \eqref{15}. The Ricci tensor corresponding to the line-element \eqref{17} can be written as of the form Eq.\eqref{6}, where the vector field $J$ is given by
\begin{equation}\label{19}
  J= -\frac{2 \left| \zeta \right| \nu ^{2} \mathcal{L}^{-}}{(a^{+}+a^{-}) ( \mathcal{L}^{+}+\mathcal{L}^{-})} \partial_{v}+\frac{2 \nu}{( \mathcal{L}^{+}+\mathcal{L}^{-})} \partial_{\phi}
\end{equation}
 Since this vector field is normalized to $l^{2}$ and satisfies Eq.\eqref{5}, thus the line-element \eqref{17} is a solution of Generalized minimal massive gravity.\\
The variation generated by the following Killing vector field preserves the form of the line-element \eqref{17}
\begin{equation}\label{20}
      \xi ^{v} = \frac{\left| \zeta \right| \nu \left( \mathcal{L}^{+} \mathcal{Z}^{-}+\mathcal{L}^{-} \mathcal{Z}^{+}\right)}{(a^{+}+a^{-}) \left( \mathcal{L}^{+}+\mathcal{L}^{-} \right)}, \hspace{0.7 cm}\xi ^{\rho} =0,\hspace{0.7 cm} \xi ^{\phi} = -\frac{\mathcal{Z}^{+}-\mathcal{Z}^{-}}{ \mathcal{L}^{+}+\mathcal{L}^{-}},
\end{equation}
where $ \mathcal{Z}^{\pm} =\mathcal{Z}^{\pm} (\phi) $ are two arbitrary periodic functions of $\phi$. In other words, under transformation generated by the Killing vector field \eqref{20}, the metric \eqref{17} transforms as
\begin{equation}\label{21}
  g_{\mu \nu}[\mathcal{L}_{+},\mathcal{L}_{-}] \rightarrow g_{\mu \nu}[\mathcal{L}_{+} + \delta _{\xi} \mathcal{L}_{+} ,\mathcal{L}_{-}+ \delta _{\xi} \mathcal{L}_{-}],
\end{equation}
with
\begin{equation}\label{22}
  \delta _{\xi} \mathcal{L}_{\pm} = \mp \partial_{\phi} \mathcal{Z}^{\pm}.
\end{equation}
Since $\xi$ depends on dynamical fields so we need to introduce a modified version of the Lie brackets \cite{8a}
\begin{equation}\label{23}
  \left[ \xi_{1} , \xi_{2} \right] =  \pounds _{\xi_{1}} \xi_{2} - \delta ^{(g)}_{\xi _{1}} \xi_{2} + \delta ^{(g)}_{\xi _{2}} \xi_{1},
\end{equation}
where $\xi_{1}= \xi(\mathcal{Z}^{+}_{1}, \mathcal{Z}^{-}_{1})$ and $\xi_{2}= \xi(\mathcal{Z}^{+}_{2}, \mathcal{Z}^{-}_{2})$. In the equation \eqref{23}, $\delta ^{(g)}_{\xi _{1}} \xi_{2}$ denotes the change induced in $\xi_{2}$ due to the variation of metric $\hat{\delta} _{_{\xi _{1}}} g_{\mu\nu} = \pounds _{\xi_{1}} g_{\mu\nu}$. By substituting Eq.\eqref{20} into Eq.\eqref{23}, we can find that
\begin{equation}\label{24}
  \left[ \xi(\mathcal{Z}^{+}_{1}, \mathcal{Z}^{-}_{1}) , \xi(\mathcal{Z}^{+}_{2}, \mathcal{Z}^{-}_{2}) \right] = 0.
\end{equation}
Thus the Killing vectors $\xi_{1}= \xi(\mathcal{Z}^{+}_{1}, \mathcal{Z}^{-}_{1})$ and $\xi_{2}= \xi(\mathcal{Z}^{+}_{2}, \mathcal{Z}^{-}_{2})$ commute and hence the algebra of the Killing vector fields is closed.
\section{Conserved charges in Generalized minimal massive gravity}\label{S.V}
In this section, we summarize some results of the Ref.\cite{7a}. The reader can refer to the Ref.\cite{9a} for a comprehensive review of the content of this section.
\subsection{Quasi-local conserved charges in Chern-Simons-like theories of gravity}\label{S.V.A}
The Lagrangian 3-form of the Chern-Simons-like theories of gravity is given by \cite{1}
\begin{equation}\label{25}
  \textbf{L}=\frac{1}{2} \textbf{g}_{rs} a^{r} \cdot da^{s}+\frac{1}{6} \textbf{f}_{rst} a^{r} \cdot a^{s} \times a^{t}.
\end{equation}
In the above Lagrangian $ a^{ra}=a^{ra}_{\hspace{3 mm} \mu} dx^{\mu} $ are Lorentz vector valued one-forms where, $r$ and $a$ indices refer to flavour and Lorentz indices, respectively. Here the wedge products of Lorentz-vector valued one-form fields are implicit. Also, $\textbf{g}_{rs}$ is a symmetric constant metric on the flavour space and $\textbf{f}_{rst}$ is a totally symmetric "flavour tensor" which are interpreted as the coupling constants. We use a 3D-vector algebra notation for Lorentz vectors in which contractions with $\eta _{ab}$ and $\varepsilon ^{abc}$ are denoted by dots and crosses, respectively \footnote{Here we consider the notation used in \cite{1}.}. It is worth saying that $a^{ra}$ is a collection of the dreibein $e^{a}$, the dualized spin-connection $\omega ^{a}$, the auxiliary field $ h^{a}_{\hspace{1.5 mm} \mu} = e^{a}_{\hspace{1.5 mm} \nu} h^{\nu}_{\hspace{1.5 mm} \mu} $ and so on \footnote{That is $a^{r}=\{ e,\omega, h, \cdots \}$, for instance, for $r=e$ and $r=\omega$ we have $a^{e}=e$ and $a^{\omega}=\omega$.}. Also for all interesting CSLTG we have $\textbf{f}_{\omega rs} = \textbf{g}_{rs}$ \cite{11a}.\\
The total variation of $a^{ra}$ induced by a diffeomorphism generator $\xi$ is \cite{12a}
\begin{equation}\label{26}
  \delta _{\xi} a^{ra} = \mathfrak{L}_{\xi} a^{ra} -\delta ^{r} _{\omega} d \chi _{\xi} ^{a} ,
\end{equation}
where $\chi _{\xi} ^{a}= \frac{1}{2} \varepsilon ^{a} _{\hspace{1.5 mm} bc} \lambda _{\xi}^{bc} $ and $\lambda _{\xi}^{bc}$ is generator of the Lorentz gauge transformations $SO(2, 1)$. Also, $ \delta ^{r} _{s} $  denotes the ordinary Kronecker delta and the Lorentz-Lie derivative along a vector field $\xi$ is denoted by $\mathfrak{L}_{\xi}$. The Lorentz-Lie derivative of a Lorentz tensor-valued $p$-form $\mathcal{A}^{a \cdots}_{b \cdots}$ is defined by
\begin{equation}\label{27}
  \mathfrak{L}_{\xi} \mathcal{A}^{a \cdots}_{b \cdots}= \pounds_{\xi} \mathcal{A}^{a \cdots}_{b \cdots} + \lambda^{\hspace{1 mm} a}_{\xi \hspace{1 mm} c} \mathcal{A}^{c \cdots}_{b \cdots}+ \cdots - \lambda^{\hspace{1 mm} c}_{\xi \hspace{1 mm} b}\mathcal{A}^{a \cdots}_{c \cdots} - \cdots.
\end{equation}
Dreibein and spin-connection \footnote{Spin-connection $\omega^{ab}$ and dualized spin-connection $\omega^{a}$ are related as $\omega^{a}= \frac{1}{2}\varepsilon^{abc}\omega_{bc}$.} transform like $e \rightarrow \Lambda e$ and $\omega \rightarrow \Lambda \omega \Lambda^{-1} + \Lambda d \Lambda^{-1}$ under Lorentz gauge transformations, where $\Lambda= \exp (\lambda) \in SO(2,1)$. Total variation induced by a vector field $\xi$ is a combination of variations due to a diffeomorphism and an infinitesimal Lorentz gauge transformation \cite{13a}. It is obvious that the total variation of a Lorentz tensor-valued $p$-form is equal to its Lorentz-Lie derivative and the extra term in total variation of $a^{ra}$ comes from the transformation law
of spin-connection under Lorentz gauge transformations. Total variation of $a^{ra}$ is covariant under Lorentz gauge transformations as well as diffeomorphism. Hence we are allowed to use covariant phase space method to obtain conserved charges in CSLTG. The arbitrary variation of the Lagrangian \eqref{25} is
\begin{equation}\label{28}
  \delta \textbf{L} = \delta a^{r} \cdot E_{r} + d \Theta (a, \delta a),
\end{equation}
with
\begin{equation}\label{29}
   E_{r}^{\hspace{1.5 mm} a} = \textbf{g}_{rs} d a^{sa} + \frac{1}{2} \textbf{f}_{rst} (a^{s} \times a^{t})^{a} , \hspace{0.7 cm} \Theta (a, \delta a) = \frac{1}{2} \textbf{g}_{rs} \delta a^{r} \cdot a^{s}.
\end{equation}
where $ E_{r}^{\hspace{1.5 mm} a} =0$ are the equations of motion and $\Theta (a, \delta a)$ is surface term. The total variation of the Lagrangian induced by diffeomorphism generator $\xi$ can be written as \footnote{$i_{\xi}$ denotes interior product in $\xi$.}
\begin{equation}\label{30}
  \delta_{\xi} \textbf{L} = \mathfrak{L}_{\xi} \textbf{L} + d \psi_{\xi}=d \left( i_{\xi} \textbf{L} + \psi_{\xi} \right),
\end{equation}
with $\psi _{\xi} = \frac{1}{2} \textbf{g}_{\omega r} d \chi_{\xi} \cdot a^{r}$. Also, the total variation of the surface term is
\begin{equation}\label{31}
  \delta_{\xi} \Theta (a, \delta a) = \mathfrak{L}_{\xi} \Theta (a, \delta a) + \Pi_{\xi},
\end{equation}
with $\Pi _{\xi}=\frac{1}{2} \textbf{g}_{\omega r} d \chi_{\xi} \cdot \delta a^{r}$. Now we assume that the variation of Lagrangian \eqref{25} is generated by a vector field $\xi$. For generality, we assume that vector field $\xi$ depends on dynamical fields. By using the Bianchi identities, we find off-shell Noether current
\begin{equation}\label{32}
  \textbf{J}_{\xi}= \Theta (a, \delta_{\xi} a) - i_{\xi} \textbf{L} - \psi_{\xi} + i_{\xi} a^{r} \cdot E_{r} - \chi _{\xi} \cdot E_{\omega}
\end{equation}
which is conserved off-shell, i.e. we have $d \textbf{J}_{\xi}=0$ off-shell. By virtue of the Poincare lemma, one can obtain off-shell Noether charge density $\textbf{K}_{\xi}$ so that $\textbf{J}_{\xi}= d \textbf{K}_{\xi}$. By taking an arbitrary variation from Eq.\eqref{32} and making some calculations one can define extended off-shell ADT \footnote{Where ADT stands for Abbott, Deser and Tekin.} current as \cite{3}
\begin{equation}\label{33}
   \mathcal{J}_{\text{ADT}} (a , \delta a, \delta _{\xi} a)= \delta a^{r} \cdot i_{\xi} E_{r} + i_{\xi} a^{r} \cdot \delta E_{r} - \chi _{\xi} \cdot \delta E_{\omega} + \textbf{g} _{rs} \delta _{\xi} a^{r} \cdot \delta a^{s}.
\end{equation}
If we assume that $\xi$ is a Killing vector field then the last term in Eq.\eqref{33} vanishes and extended off-shell ADT current reduces to the generalized off-shell ADT current \cite{12a}. Also, if we consider on-shell case, i.e. $E_{r}=\delta E_{r}=0$, then extended off-shell ADT current reduces to symplectic current \cite{15a,16a,17a}. The current \eqref{33} is conserved off-shell, i.e. $d\mathcal{J}_{\text{ADT}}=0$, so by virtue of the Poincare lemma, one can find corresponding extended off-shell ADT charge so that $\mathcal{J}_{\text{ADT}}=d \mathcal{Q}_{\text{ADT}}$. Therefore we can define quasi-local conserved charge perturbation associated with a field dependent vector field $\xi$ as
\begin{equation}\label{34}
\begin{split}
   \delta Q ( \xi )  & =\frac{1}{8 \pi G} \int_{\Sigma} \mathcal{Q}_{\text{ADT}} \\
     & =\frac{1}{8 \pi G} \int_{\Sigma} \left( \textbf{g}_{rs} i_{\xi} a^{r} - \textbf{g} _{\omega s} \chi _{\xi} \right) \cdot \delta a^{s},
\end{split}
\end{equation}
where $G$ denotes the Newtonian gravitational constant and $\Sigma$ is a spacelike codimension two surface. We can take an integration from \eqref{34} over the one-parameter path on the solution space \cite{18a,19a} and then we find that
\begin{equation}\label{35}
  Q ( \xi )  = \frac{1}{8 \pi G} \int_{0}^{1} ds \int_{\Sigma} \left( \textbf{g}_{rs} i_{\xi} a^{r} - \textbf{g} _{\omega s} \chi _{\xi} \right) \cdot \delta a^{s},
\end{equation}
Also, we argued that the quasi-local conserved charge \eqref{35} is not only conserved for the Killing vectors which are admitted by spacetime everywhere but also it is conserved for the asymptotically Killing vectors. To calculate the conserved charge using Eq.\eqref{35}, we need an explicit expression for $\chi _{\xi}$ in an appropriate manner. To this end, the authors in \cite{20a} demand that it must be chosen so that the Lorentz-Lie derivative of dreibein vanishes when $\xi$ is a Killing vector field and then $\chi _{\xi}$ should be provided as follows \cite{13a}:
\begin{equation}\label{36}
  \chi _{\xi} ^{a} = i_{\xi} \omega ^{a} + \frac{1}{2} \varepsilon ^{a}_{\hspace{1.5 mm} bc} e^{\nu b} (i_{\xi} T^{c})_{\nu} + \frac{1}{2} \varepsilon ^{a}_{\hspace{1.5 mm} bc} e^{b \mu} e^{c \nu} \nabla _{\mu} \xi _{\nu} .
\end{equation}
Now we are ready to find conserved charges in the context of Generalized minimal massive gravity.
\section{Generalized Minimal Massive Gravity}\label{S.V.B}
Generalized minimal massive gravity is an example of the Chern-Simons-like theories of gravity \cite{5'}. In this model, there are four flavours of one-form, $a^{r}= \{ e, \omega , h, k \}$, and the non-zero components of the flavour metric and the flavour tensor are
\begin{equation}\label{37}
\begin{split}
     & \textbf{g}_{e \omega}=-\sigma, \hspace{1 cm} \textbf{g}_{e h}=1, \hspace{1 cm} \textbf{g}_{\omega k}=-\frac{1}{m^{2}}, \hspace{1 cm} \textbf{g}_{\omega \omega}=\frac{1}{\mu}, \\
     & \textbf{f}_{e \omega \omega}=-\sigma, \hspace{1 cm} \textbf{f}_{e h \omega}=1, \hspace{1 cm} \textbf{f}_{k \omega \omega}=-\frac{1}{m^{2}}, \hspace{1 cm} \textbf{f}_{\omega \omega \omega}=\frac{1}{\mu},\\
     & \textbf{f} _{ekk}= -\frac{1}{m^{2}}, \hspace{1 cm} \textbf{f}_{eee}=\Lambda_{0},\hspace{1 cm} \textbf{f}_{ehh}= \alpha .
\end{split}
\end{equation}
where $\sigma$, $\Lambda _{0}$, $\mu$, $m$ and $\alpha$ are a sign, cosmological parameter with dimension of mass squared, mass parameter of Lorentz Chern-Simons term, mass parameter of New Massive Gravity term and a dimensionless parameter, respectively. The equations of motion of Generalized minimal massive gravity  are \cite{5'} (see also \cite{140})
\begin{equation}\label{38}
   - \sigma R (\Omega) + (1 + \sigma \alpha ) D(\Omega) h - \frac{1}{2} \alpha (1 + \sigma \alpha ) h \times h + \frac{\Lambda _{0}}{2} e \times e - \frac{1}{2 m^{2}} k \times k  =0,
\end{equation}
\begin{equation}\label{39}
  - e \times k + \mu (1 + \sigma \alpha ) e \times h - \frac{\mu}{m^{2}} D(\Omega) k + \frac{\mu \alpha}{m^{2}} h \times k=0 ,
\end{equation}
\begin{equation}\label{40}
  R(\Omega) - \alpha D(\Omega) h + \frac{1}{2} \alpha ^{2} h \times h + e \times k =0,
\end{equation}
\begin{equation}\label{41}
  T(\Omega) = 0 ,
\end{equation}
where
\begin{equation}\label{42}
  \Omega = \omega - \alpha h
\end{equation}
is ordinary torsion-free dualized spin-connection. Also, $R(\Omega) = d \Omega + \frac{1}{2} \Omega \times \Omega$ is curvature 2-form, $T(\Omega)= D(\Omega)e$ is torsion 2-form, and $ D(\Omega) $ denotes exterior covariant derivative with respect to torsion-free dualized spin-connection.\\
Now, consider a metric so that Ricci tensor corresponding to the given metric can be written as Eq.\eqref{6} in which the vector field $J$ is normalized to $l^{2}$ and satisfies Eq.\eqref{5} (metrics \eqref{1}, \eqref{11}, \eqref{15} and \eqref{17} are examples of such a case). The authors in Ref.\cite{7a} showed that such a metric is a solution of Generalized minimal massive gravity provided that the following equations are held
\begin{equation}\label{43}
     h^{a}_{\hspace{1.5 mm \mu}}= H_{1} e^{a}_{\hspace{1.5 mm \mu}} +H_{2} J^{a}J_{\mu}, \hspace{0.7 cm} k^{a}_{\hspace{1.5 mm \mu}}= F_{1} e^{a}_{\hspace{1.5 mm \mu}} +F_{2} J^{a}J_{\mu},
\end{equation}
\begin{equation}\label{44}
  \frac{\zeta^{2}}{4l^{2}} - \frac{1}{2} \alpha l \left| \zeta \right| H_{2} - \alpha ^{2} H_{1} \left( H_{1} + l^{2} H_{2} \right) - \left( 2F_{1}+l^{2} F_{2} \right)=0,
\end{equation}
\begin{equation}\label{45}
  -\frac{\zeta^{2}}{l^{4}} \left( 1-\nu^{2} \right) + \frac{3 \alpha}{2l} \left| \zeta \right| H_{2} + \alpha ^{2} H_{1} H_{2} +F_{2}=0,
\end{equation}
\begin{equation}\label{46}
      \frac{1}{\mu} \left( 2F_{1}+l^{2} F_{2} \right)-\left( 1+ \alpha \sigma\right) \left( 2 H_{1} + l^{2} H_{2} \right) - \frac{l}{2 m^{2}} \left| \zeta \right| F_{2} - \frac{\alpha}{m^{2}} \left[ 2 H_{1}F_{1} + l^{2} \left( H_{1}F_{2}+H_{2}F_{1} \right) \right]=0,
\end{equation}
\begin{equation}\label{47}
  -\frac{1}{\mu} F_{2} + \left( 1+ \alpha \sigma\right) H_{2} + \frac{3}{2lm^{2}} \left| \zeta \right| F_{2} +\frac{\alpha}{m^{2}} \left( H_{1}F_{2}+H_{2}F_{1} \right) =0,
\end{equation}
\begin{equation}\label{48}
  - \frac{\zeta^{2}}{4l^{2}} \sigma + \frac{1}{2} \left( 1+ \alpha \sigma\right) l \left| \zeta \right| H_{2} + \alpha \left( 1+ \alpha \sigma\right) H_{1} \left( H_{1} + l^{2} H_{2} \right) - \Lambda _{0} + \frac{1}{m^{2}} F_{1} \left( F_{1}+l^{2} F_{2} \right)=0,
\end{equation}
\begin{equation}\label{49}
  \frac{\zeta^{2}}{l^{4}} \left( 1-\nu^{2} \right) \sigma -\frac{3}{2l} \left( 1+ \alpha \sigma\right) \left| \zeta \right| H_{2}- \alpha \left( 1+ \alpha \sigma\right) H_{1}H_{2} - \frac{1}{m^{2}} F_{1} F_{2}=0,
\end{equation}
where $H_{1}$, $H_{2}$, $F_{1}$, $F_{2}$ are constant parameters and $J^{a} = e^{a}_{\hspace{1.5 mm \mu}} J^{\mu} $ (for arriving at these equations appendix B in \cite{9a} can be helpful). In fact Eqs.\eqref{43} are ansatz for auxiliary fields $h$ and $k$ and equations \eqref{44}-\eqref{49} are conditions on constant parameters.\\
Using equations \eqref{37}, \eqref{42}-\eqref{49}, one can simplify the conserved charge perturbation \eqref{34} for given metrics in the context of the Generalized minimal massive gravity
\begin{equation}\label{50}
  \begin{split}
      \delta Q(\xi) = \frac{1}{8\pi G} \int_{\Sigma}\biggl\{& - \left( \sigma + \frac{\alpha H_{1}}{\mu} + \frac{F_{1}}{m^{2}}\right) \left[ i_{\xi} e \cdot \delta \Omega + \left( i_{\xi} \Omega - \chi _{\xi}\right)\cdot \delta e \right]  \\
       & +\frac{1}{\mu} \left( i_{\xi} \Omega - \chi _{\xi}\right) \cdot \delta \Omega + \alpha H_{2} \left( \frac{\alpha H_{2}}{\mu} + \frac{2 F_{2}}{m^{2}}\right) i_{\xi} \mathfrak{J} \cdot \delta \mathfrak{J} \\
       & +\left[ - \frac{\zeta^{2}}{\mu l^{2}} \left( \frac{3}{4} - \nu^{2} \right) + l \left| \zeta \right| \left( \frac{\alpha H_{2}}{\mu} + \frac{F_{2}}{m^{2}}\right) \right] i_{\xi} e \cdot \delta e \\
       & -\left( \frac{\alpha H_{2}}{\mu} + \frac{F_{2}}{m^{2}}\right) \left[ i_{\xi} \mathfrak{J} \cdot \delta \Omega + \left( i_{\xi} \Omega - \chi _{\xi}\right)\cdot \delta \mathfrak{J} \right] \\
       & +\left[ \frac{\zeta^{2}}{\mu l^{4}} \left( 1 - \nu^{2} \right) - \frac{3\left| \zeta \right|}{2l} \left( \frac{\alpha H_{2}}{\mu} + \frac{F_{2}}{m^{2}}\right) \right] \left( i_{\xi} \mathfrak{J} \cdot \delta e + i_{\xi}e \cdot \delta \mathfrak{J} \right) \biggr\}.
  \end{split}
\end{equation}
where $ \mathfrak{J}^{a}_{\hspace{1.5 mm} \mu}=J^{a}J_{\mu} $ for simplicity. One can show that the expression \eqref{36} for $\chi_{\xi}$ can be rewritten as \cite{18a}
\begin{equation}\label{51}
  i_{\xi} \Omega - \chi _{\xi} = - \frac{1}{2} \varepsilon ^{a}_{\hspace{1.5 mm} bc} e^{b \mu} e^{c \nu} \nabla _{\mu} \xi _{\nu} .
\end{equation}
Also we mention that the torsion free spin-connection is given by
\begin{equation}\label{52}
  \Omega ^{a} _{ \hspace{1.5 mm}\mu} = \frac{1}{2} \varepsilon^{a b c} e _{b} ^{ \hspace{1.5 mm} \alpha} \nabla _{\mu} e_{c \alpha}.
\end{equation}
Now we are ready to find conserved charges of warped black flower described by metric \eqref{17} associated with Killing vector field \eqref{20}.
\section{Conserved charges related to new fall-off conditions in Generalized minimal massive gravity}\label{S.VI}
Since the relation between metric tensor and dreibein is given by $g_{\mu \nu}= \eta_{ab}e^{a}_{\hspace{1.5 mm} \mu} e^{b}_{\hspace{1.5 mm} \nu} $ then, one can check that dreibein corresponding to the line-element \eqref{17} is
\begin{equation}\label{53}
  \begin{split}
     e^{0}= &\frac{1}{2} \left[ l^{2}-(a^{+}+a^{-}) \rho\right] dv - d \rho \\
            &+ \frac{1}{2} \left| \zeta \right| \nu^{-1} \left\{ \nu^{2} \left[ l^{2}-(a^{+}+a^{-}) \rho\right] \left( \frac{L^{-}}{a^{+}+a^{-}} \right)+ \rho \left[ 1+ \sqrt{1-\nu^{2}} \right] (L^{+}+L^{-}) \right\} d \phi \\
     e^{1}=  & l \left| \zeta \right| \rho \sqrt{1-\nu^{2}} dv + l \nu^{-1} \left[ \frac{1}{2} (L^{+}+L^{-}) + \frac{\zeta^{2} \nu^{2} L^{-} \rho}{(a^{+}+a^{-})} \sqrt{1-\nu^{2}} \right] d \phi \\
      e^{2}= & \frac{1}{2} \left[ l^{2}+(a^{+}+a^{-}) \rho\right] dv + d \rho \\
            &+ \frac{1}{2} \left| \zeta \right| \nu^{-1} \left\{ \nu^{2} \left[ l^{2}+(a^{+}+a^{-}) \rho\right] \left( \frac{L^{-}}{a^{+}+a^{-}} \right)- \rho \left[ 1+ \sqrt{1-\nu^{2}} \right] (L^{+}+L^{-}) \right\} d \phi
  \end{split}
\end{equation}
Now, we take the spacelike codimension two surface $\Sigma$ in Eq.\eqref{50} to be a circle with radius $\rho \rightarrow 0$. By substituting Eq.\eqref{53} and Eq.\eqref{20} into Eq.\eqref{50} and using Eqs.\eqref{44}-\eqref{49}, \eqref{51} and \eqref{52}, we find the following expression for conserved charge associated with the Killing vector field \eqref{20}
\begin{equation}\label{54}
  Q(\xi)= Q^{+}(\mathcal{Z}^{+})+Q^{-}(\mathcal{Z}^{-}),
\end{equation}
with
\begin{equation}\label{55}
  Q^{\pm}(\mathcal{Z}^{\pm})= \frac{\zeta^{2} \nu^{2} c_{\pm}}{96 \pi} \int_{0}^{2 \pi} \mathcal{Z}^{\pm}(\phi) \mathcal{L}^{\pm}(\phi) d \phi ,
\end{equation}
where
\begin{equation}\label{56}
  c_{+}=\frac{3 l }{\left| \zeta \right| \nu^{2} G} \left\{ \sigma + \frac{\alpha}{\mu} \left(H_{1}+l^{2}H_{2} \right)+\frac{1}{m^{2}} \left(F_{1}+l^{2}F_{2} \right)-\frac{\left| \zeta \right|}{2\mu l}\right\},
\end{equation}
\begin{equation}\label{57}
  c_{-}=\frac{3 l }{\left| \zeta \right| \nu^{2} G}\left\{ \sigma + \frac{\alpha}{\mu} \left(H_{1}+l^{2}H_{2} \right)+\frac{1}{m^{2}} \left(F_{1}+l^{2}F_{2} \right)-\frac{\left| \zeta \right|}{2\mu l} \left( 1-2\nu^{2}\right)\right\}.
\end{equation}
To interpret $c_{\pm}$, we deviate slightly from the discussion. The authors in the paper \cite{7a} showed that the algebra of asymptotic conserved charges of asymptotically (at spatial infinity) spacelike warped AdS$_{3}$ spacetimes in Generalized minimal massive gravity is given as the semi direct product of the Virasoro algebra with U(1) current algebra, with central charges $c_{V}= c_{-}$ and $c_{U}= \zeta^{-2} \nu^{-4} c_{+}$. The algebra of these conserved charges does not describe the conformal symmetry. Therefore, they used a particular Sugawara construction to reconstruct the conformal algebra. Strictly speaking, they mapped the algebra to two Virasoro algebras with central charges $c_{\pm}$.\\
Now we return to our discussion. The algebra of conserved charges can be written as \cite{12,24a}
\begin{equation}\label{58}
  \left\{ Q(\xi _{1}) , Q(\xi _{2}) \right\} = Q \left(  \left[ \xi _{1} , \xi _{2} \right] \right) + \mathcal{C} \left( \xi _{1} , \xi _{2} \right)
\end{equation}
where $\mathcal{C} \left( \xi _{1} , \xi _{2} \right)$ is central extension term. Also, the left hand side of the equation \eqref{53} can be defined by
\begin{equation}\label{59}
  \left\{ Q(\xi _{1}) , Q(\xi _{2}) \right\}= \delta _{\xi _{2}} Q(\xi _{1}).
\end{equation}
Therefore, because of the Eq.\eqref{24} one can deduce that the central extension term can be found as follows:
\begin{equation}\label{60}
\begin{split}
   \mathcal{C} \left( \xi _{1} , \xi _{2} \right)= & \delta _{\xi _{2}} Q(\xi _{1}) \\
    = & \frac{\zeta^{2} \nu^{2}}{96 \pi} \int_{0}^{2 \pi} \left(- c_{+} \mathcal{Z}^{+}_{1} \partial_{\phi} \mathcal{Z}^{+}_{2}+ c_{-} \mathcal{Z}^{-}_{1} \partial_{\phi} \mathcal{Z}^{-}_{2} \right) d \phi.
\end{split}
\end{equation}
In this way, the algebra of conserved charges can be written as
\begin{equation}\label{61}
  \left\{ Q^{\pm}(\mathcal{Z}^{\pm}_{1}) , Q^{\pm}(\mathcal{Z}^{\pm}_{2}) \right\} =\mp \frac{\zeta^{2} \nu^{2} c_{\pm}}{96 \pi} \int_{0}^{2 \pi} \mathcal{Z}^{\pm}_{1} \partial_{\phi} \mathcal{Z}^{\pm}_{2} d \phi, \hspace{0.7 cm} \left\{ Q^{+}(\mathcal{Z}^{+}_{1}) , Q^{-}(\mathcal{Z}^{-}_{2}) \right\}=0.
\end{equation}
By setting $\mathcal{Z}^{\pm}=e^{in \phi}$, one can expand $Q^{\pm}(\mathcal{Z}^{\pm})$ in Fourier modes
\begin{equation}\label{62}
  \mathcal{L}^{\pm}_{n}= \frac{\zeta^{2} \nu^{2} c_{\pm}}{96 \pi} \int_{0}^{2 \pi} e^{in \phi} \mathcal{L}^{\pm}(\phi) d \phi .
\end{equation}
Also, by substituting $\mathcal{Z}^{\pm}_{1}=e^{in \phi}$ and $\mathcal{Z}^{\pm}_{2}=e^{im \phi}$ into Eq.\eqref{61} and replacement of Dirac brackets by commutators $i \{ , \} \rightarrow [,]$, we have
\begin{equation}\label{63}
  \left[ \hat{\mathcal{L}}^{\pm}_{n} , \hat{\mathcal{L}}^{\pm}_{m} \right] =\mp \frac{\zeta^{2} \nu^{2} c_{\pm}}{48} n \delta_{m+n,0}, \hspace{0.7 cm} \left[ \hat{\mathcal{L}}^{+}_{n} , \hat{\mathcal{L}}^{-}_{m} \right]=0.
\end{equation}
Similar to the near horizon symmetry algebra of the black flower solutions in Generalized minimal massive gravity \cite{3}, the near horizon symmetry algebra of the warped black flower consists of two $U(1)$ current algebras, with different levels. In this case, the levels are $\mp \frac{\zeta^{2} \nu^{2} c_{\pm}}{48}$. One can change the basis according to following definitions
\begin{equation}\label{64}
\begin{split}
   \hat{X}_{n} = & -\frac{i}{\left| \zeta\right| \nu} \left( \sqrt{\frac{24}{c_{+}}} \hat{\mathcal{L}}_{n}^{+} + \sqrt{\frac{24}{c_{-}}} \hat{\mathcal{L}}_{n}^{-} \right) \hspace{0.5 cm}  \text{for} \hspace{0.3 cm} n \in \mathbb{Z} \\
   \hat{P}_{n} = & \frac{1}{\left| \zeta\right| \nu n} \left( \sqrt{\frac{24}{c_{+}}} \hat{\mathcal{L}}_{-n}^{+} -  \sqrt{\frac{24}{c_{-}}} \hat{\mathcal{L}}_{-n}^{-} \right) \hspace{0.5 cm}  \text{for} \hspace{0.3 cm} n \neq 0 \\
    \hat{P}_{0} = & \hat{\mathcal{L}}_{0}^{+} + \hat{\mathcal{L}}_{0}^{-} \hspace{0.5 cm}  \text{for} \hspace{0.3 cm} n = 0,
\end{split}
\end{equation}
By using  the above equations, the algebra \eqref{63}, takes following form
\begin{equation}\label{65}
  \left[ \hat{X}_{n} , \hat{X}_{m} \right] = \left[ \hat{P}_{n} , \hat{P}_{m} \right] = \left[ \hat{X}_{0} , \hat{P}_{n} \right] = \left[ \hat{P}_{0} , \hat{X}_{n} \right] =0
\end{equation}
\begin{equation}\label{66}
  \left[ \hat{X}_{n} , \hat{P}_{m} \right] = i \delta _{nm} \hspace{0.5 cm} \text{for} \hspace{0.3 cm} n,m \neq 0.
\end{equation}
It is clear that $ \hat{X}_{0}$ and $ \hat{P}_{0}$ are the two Casimirs and Eq.\eqref{66} is the Heisenberg algebra. It is interesting that, in the given model also we obtain the Heisenberg algebra as the near horizon symmetry algebra of the warped black flower solutions. By comparing the definition of $\hat{P}_{0}$ and Eq.\eqref{54}, one can deduce that $\hat{P}_{0}$ is just the Hamiltonian, i.e. $\hat{H} \equiv \hat{P}_{0}$.
\section{Soft hair and the soft hairy warped black hole entropy}\label{S.VII}
The Hamiltonian $\hat{H} \equiv \hat{P}_{0}$ gives us the dynamics of the system near the horizon. Let us consider all vacuum descendants \cite{9}
\begin{equation}\label{67}
  | \psi (q) \rangle = N(q) \prod _{i=1} ^{N^{+}} \left( \hat{\mathcal{L}}_{-n_{i}^{+}}^{+} \right)^{m_{i}^{+}} \prod _{i=1} ^{N^{-}} \left( \hat{\mathcal{L}}_{-n_{i}^{-}}^{-} \right)^{m_{i}^{-}} | 0 \rangle
\end{equation}
where $q$ is a set of arbitrary non-negative integer quantum numbers $N^{\pm}$, $n_{i}^{\pm}$ and $m_{i}^{\pm}$. Also, $N(q)$ is a normalization constant such that $\langle \psi (q) | \psi (q) \rangle = 1$. The Hamiltonian $\hat{H} \equiv \hat{P}_{0}=\hat{\mathcal{L}}_{0}^{+} + \hat{\mathcal{L}}_{0}^{-}$ commutes with all generators $\hat{\mathcal{L}}_{n}^{\pm}$, so the energy of all states are the same. The energy of the vacuum state is given by the following eigenvalue equation
\begin{equation}\label{68}
  \hat{H}  | 0 \rangle = E_{\text{vac}}  | 0 \rangle .
\end{equation}
Also, for all descendants, we have
\begin{equation}\label{69}
  E_{\psi} = \langle \psi (q) | H  | \psi (q) \rangle .
\end{equation}
Due to the mentioned property of the Hamiltonian, we find that all descendants of the vacuum have the same energy as the vacuum,
\begin{equation}\label{70}
  E_{\psi} = E_{\text{vac}},
\end{equation}
in other words, they are soft hairs in the sense of being zero-energy excitations \cite{9,26a}.\\
For the case of the warped black hole, we have
\begin{equation}\label{71}
  \mathcal{L}^{+}=  r_{+} + r_{-} +2 \nu \sqrt{r_{+} r_{-}}, \hspace{0.7 cm} \mathcal{L}^{-}=  r_{+} - r_{-} , \hspace{0.7 cm} a^{\pm} =- \frac{\left| \zeta \right| \nu^{2}\left( r_{+}-r_{-} \right)}{2\left( r_{+} + \nu \sqrt{r_{+} r_{-}} \right)},
\end{equation}
By substituting Eq.\eqref{71} into Eq.\eqref{62}, we find the eigenvalues of $ \hat{\mathcal{L}}_{n}^{\pm}$ as follows:
\begin{equation}\label{72}
  \mathcal{L}_{n}^{+} = \frac{\zeta^{2} \nu^{2} c_{+}}{48} (r_{+} + r_{-} +2 \nu \sqrt{r_{+} r_{-}}) \delta _{n,0}, \hspace{0.7 cm}\mathcal{L}_{n}^{-} = \frac{\zeta^{2} \nu^{2} c_{-}}{48} (r_{+} - r_{-}) \delta _{n,0} .
\end{equation}
The entropy of a soft hairy black hole is related to the zero mode charges $\mathcal{L}_{0}^{\pm}$ by the following formula \cite{9,26a,27a,28a,29a,30a,31a}
\begin{equation}\label{73}
  \mathcal{S}=2\pi \left( \mathcal{L}_{0}^{+} + \mathcal{L}_{0}^{-} \right).
\end{equation}
Hence, by substituting Eq.\eqref{72} into Eq.\eqref{73}, we find the entropy of the warped black hole solution of Generalized minimal massive gravity as
\begin{equation}\label{74}
\begin{split}
   \mathcal{S}= & \frac{\pi l \left| \zeta \right|}{4G} \biggl\{ \left[ \sigma + \frac{\alpha}{\mu} \left(H_{1}+l^{2}H_{2} \right)+\frac{1}{m^{2}} \left(F_{1}+l^{2}F_{2} \right)-\frac{\left| \zeta \right|}{2\mu l} \right] \left( r_{+} + \nu \sqrt{r_{+} r_{-}} \right)\\
     & + \frac{\left| \zeta \right| \nu^{2}}{2 \mu l} \left( r_{+}-r_{-}\right) \biggr\}.
\end{split}
\end{equation}
which is exactly matched with the results of the paper \cite{7a}. As we know, $\hat{\mathcal{L}}_{0}^{+} + \hat{\mathcal{L}}_{0}^{-}= \hat{P}_{0}$  is one of two Casimirs of algebra, i.e. $\hat{P}_{0}$ is a constant of motion. Therefore, one expects that the zero mode eigenvalue of $\hat{P}_{0}$ should be correspond to a conserved charge of given spacetime. We have shown that entropy is the intended conserved charge in the context of Generalized minimal massive gravity.
\section{Conclusion}
We have considered spacelike warped AdS$_{3}$ black hole metric in Boyer-Lindquist coordinate system (see Eq.\eqref{1}). The spacetime described by metric \eqref{1} admits $SL(2,\mathbb{R})\times U(1)$ as isometry group. Then the Ricci tensor corresponding to that metric can be written as Eq.\eqref{6}, where the vector field $J$ is normalized to $l^{2}$ and satisfies Eq.\eqref{5}. We have deduced in subsection.\ref{S.V.B} that each metric with this property will be a solution of the Generalized minimal massive gravity. Hence, we tried to present boundary conditions so that they should respect to that property. The near horizon geometry of three-dimensional black holes can be expressed using Gaussian null coordinates. Then we introduced coordinates transformation \eqref{10} so that it maps metric \eqref{1} into the one in Gaussian null coordinates. Inspired by the obtained metric \eqref{11}, we introduced the boundary conditions near the horizon of non-extremal warped black holes (see Eq.\eqref{17}). The line-element \eqref{17} is an exact solution of the Generalized minimal massive gravity. The geometry possesses an event horizon located at $\rho=0$. This geometry is not spherically symmetric, but is still stationary. So this solution describes a ''warped black flower''. We obtained conserved charges of the new fall-off conditions and showed that the near horizon symmetry algebra of such solutions in the context of Generalized minimal massive gravity consists of two $U(1)$ current algebras, with the levels $\mp \frac{\zeta^{2} \nu^{2} c_{\pm}}{48}$. By changing the basis according to the definitions \eqref{64} we obtained the Heisenberg algebra as the near horizon symmetry algebra of the warped black flower solutions. We inferred that the vacuum state and all descendants of the vacuum have the same energy. Thus these zero energy excitations on the horizon appear as soft hairs on the warped black hole. Since the Hamiltonian is given by $\hat{H} \equiv \hat{P}_{0}=\hat{\mathcal{L}}_{0}^{+} + \hat{\mathcal{L}}_{0}^{-}$, and $\hat{P}_{0}$ is a Casimir of the algebra then by finding the eigenvalues of $\hat{\mathcal{L}}_{n}^{\pm}$ for the warped black hole we have checked that the formula for the entropy of a soft hairy black hole $\mathcal{S}=2\pi ( \mathcal{L}_{0}^{+} + \mathcal{L}_{0}^{-} )$ gives us the correct value of the entropy of the warped black hole solution of the Generalized minimal massive gravity. By setting $\sigma =-1$, $\alpha \rightarrow 0$ and $m^{2} \rightarrow \infty$, where the GMMG reduce to the TMG with negative cosmological constant, all our results for the GMMG, reduced to the results which one can expect for topological massive gravity \cite{1c}.
\section{Acknowledgments}
This work has been financially supported by Research Institute for Astronomy and Astrophysics of Maragha (RIAAM).

\end{document}